\input{aipcheck}

\documentclass[
    ,final            
  ]
  {aipproc}

\layoutstyle{6x9}

\def\pythia{{\sc Pythia}}
\def\cascade{{\sc Cascade}}
\def\herwig{{\sc Herwig}}
\def\disent{{\sc Disent}}
\def\mcatnlo{MC@NLO}

\begin{document}

\hskip 11 cm {\rm \small OUTP-08-11-P}

\title{Jet correlations from unintegrated parton distributions}


\author{F.~Hautmann}{
  address={Oxford University, Theoretical Physics Department, Oxford OX1 3NP}
}

\author{H.~Jung}{
  address={Deutsches Elektronen Synchrotron, D-22603 Hamburg}
}

\begin{abstract}
Transverse-momentum dependent parton distributions can be introduced 
gauge-invariantly in QCD from high-energy factorization. We discuss 
Monte Carlo applications of these distributions to parton showers and jet 
physics, with a view to the implications for the  Monte Carlo description  
 of  complex hadronic final states with multiple hard scales at the LHC. 
\end{abstract}

\maketitle


\vskip 0.5 cm 

{\em  \hspace*{ 1.2 cm}  Sixth International Conference on 
Perspectives in Hadronic Physics\\ 
\hspace*{ 4.6 cm} ICTP, Trieste, 12-16 May 2008\\
        \hspace*{ 4.7 cm}          AIP Conference Proceedings}

\vskip 0.5 cm

\section{Introduction}

Experimental studies of multi-particle final states at high-energy hadron  colliders 
rely on realistic event simulation by parton-shower Monte Carlo generators. 
Multi-particle states acquire qualitatively new features at the forthcoming 
Large Hadron Collider (LHC)   compared to previous hadron-hadron experiments 
due to the large phase space opening up for events characterized by multiple hard 
scales $ q_1^2, ..., q_n^2 $,  possibly widely disparate from each other. 
This brings in potentially large radiative corrections logarithmic in the ratio of 
two such scales, $\alpha_s^k \ln^m q_i^2 / q_j^2 $, and potentially new effects 
in the nonperturbative components of production processes (e.g., parton densities 
being probed in regions of the phase space  near the kinematic boundaries). 
It is not at all obvious that the approximations involved in standard Monte Carlo 
generators that have successfully served for event simulation in past  collider 
experiments will be up to the new situation. 

Standard parton-shower generators like 
\herwig\ and \pythia\  are based on the dominance of 
collinear QCD radiation, supplemented by  color-coherence effects 
for soft gluon emission from partons carrying longitudinal 
momentum fraction $x \sim {\cal O} (1)$. 
However as the energy increases 
the effects of emissions that are not collinearly ordered 
are known to become more and more important,  and coherence effects from 
space-like partons carrying momentum fractions $x \ll 1$ set in. 
 The high-energy  multi-scale kinematics is bound to enhance the 
sensitivity to  these dynamical features. 
The theoretical framework to take this into account 
requires the use of generalized QCD  factorization 
techniques and the introduction of partonic distributions 
unintegrated not only in the longitudinal momenta 
but also in the  transverse momenta. 
Although Monte Carlo implementations of this framework  are 
not nearly as developed as standard  shower generators at present, 
there are already studies that 
have been able to show their potential advantages 
over  collinear-based  algorithms 
in specific cases of hadronic final-state analyses. 

In this article  we briefly recall the basis for the introduction of 
transverse-momentum dependent (TMD) parton distributions from high-energy 
factorization, and  point to ongoing activity 
toward fully general definitions; then 
we  discuss Monte Carlo calculations that  use  the high-energy TMD  framework to make 
predictions for  jet observables,  including angular and momentum correlations in 
final states with multiple jets. We  comment on current developments 
toward  general-purpose Monte Carlo tools and applications to 
 final states with heavy quarks and heavy bosons  plus jets at the LHC.

\section{TMD distributions from high-energy factorization }

Precise 
definitions for transverse-momentum dependent (TMD), or unintegrated,  parton 
distribution functions~\cite{jcc-lc08,jcc03} are 
at the center of much current activity. 
In the general case, to characterize such distributions gauge-invariantly 
over the whole phase space is a difficult question, and a number of open issues remain.  
In the case of small $x$ a well-prescribed, gauge-invariant definition 
emerges from high-energy factorization~\cite{hef}, and has been used for studies 
of  collider processes both  by Monte Carlo~\cite{jeppe04,hj_rec} and by 
semi-analytic resummation~\cite{radcortalk1,radcortalk2} approaches. 

The diagrammatic argument for gauge invariance,  
 given in~\cite{hef}, and developed in~\cite{hef94}, is based on relating 
off-shell matrix elements with physical 
cross sections  at $x \ll 1$, and exploits   the dominance of  
single gluon polarization at high energies.\footnote{It 
is emphasized e.g. in~\cite{jeppe04,jccdis01} that 
a fully worked out 
 operator argument, on the other hand,  is highly desirable but is still 
missing.} 
The main reason why a natural definition for TMD 
pdfs can be constructed in the high-energy limit 
is that   one can relate directly (up to 
perturbative corrections) the cross section for a {\em physical}  
process, say, photoproduction of a heavy-quark pair, 
to an {\em unintegrated} gluon distribution, much as, in the conventional 
parton picture, one does for DIS in terms of ordinary (integrated) parton 
distributions. 
On the other hand, the difficulties 
in defining a TMD distribution over the whole phase space can 
largely be associated  with the fact that 
it is not  obvious how to determine one such 
 relation for general kinematics. 

The evolution equations obeyed by TMD distributions defined from the 
 high-energy limit are of the type of energy evolution~\cite{lip97}.  
 Factorization formulas in terms of TMD  distributions~\cite{hef}  
have corrections that are down   by logarithms of energy 
rather than powers of momentum transfer. On the other 
hand, it is important to observe that 
 this framework   allows one 
 to describe the ultraviolet region of arbitrarily high k$_\perp$   
and in particular re-obtain the structure of QCD 
logarithmic scaling violations~\cite{radcortalk1,radcortalk2,hef94}. 
  This  ultimately 
justifies the use of this approach for jet physics.       
 In particular it is the basis for  using  corresponding  
  Monte Carlo implementations~\cite{hj04} 
  to treat multi-scale hard processes at the LHC.

From both 
theoretical and phenomenological viewpoints,   
it is one of the 
appealing features  of the high-energy framework for  TMD distributions 
that one can  relate its results  to a well-defined summation of 
higher-order radiative corrections. By expanding these 
results to fixed order in $\alpha_s$, one can match  the predictions thus obtained 
against perturbative calculations. This has been verified  for  a number of specific 
processes at next-to-leading order (see for instance~\cite{vannee} for heavy flavor 
production) and more recently at next-to-next-to-leading order (see for 
instance~\cite{mochetal}). 
Note that this fact also provides the basis for  
 shower algorithms implementing this framework to be combined with 
fixed-order NLO  calculations by using existing techniques for such 
matching. 

In the next section we point to topical issues and 
activity on   TMD generalizations.  After this, 
we   focus on  existing high-energy Monte Carlo with unintegrated pdfs 
and phenomenological applications to jets.

\section{Low energies }

As mentioned above,  in the general case  full results on   
 TMD  distributions  are yet to be established. 
The current status is discussed in~\cite{jcc-lc08}.  
 Factorization formulas 
in terms of unintegrated parton distributions will have a considerably 
complex structure. 
A prototypical calculation that illustrates this structure 
 is carried out in~\cite{jccfh00}, which treats,  rather 
than a general scattering  observable,  a simpler problem,  the 
electromagnetic form factor of a quark. 
This case is however sufficient to illustrate certain 
 main features, namely the role of nonperturbative, gauge-invariantly defined 
 factors associated with infrared subgraphs (both collinear and soft), 
and the role of infrared subtractive techniques 
that serve to identify these factors.   
Analyses  along these lines  for  more general processes,  
 involving fully unintegrated pdfs,  have recently 
been reported by T.~Rogers~\cite{rogers}. 

One of the questions that a full factorization statement will  address 
is the treatment of soft gluons exchanged between 
subgraphs in different collinear directions. The underlying 
dynamics is that of  non-abelian 
Coulomb phase, treated a long time ago in~\cite{dyproof}  
for the fully inclusive Drell-Yan case.  
But a systematic treatment for more complex observables, including color 
in both initial and final states, is still  missing.    
Vogelsang and Yuan~\cite{vogel0708} illustrate this  point neatly 
with a simplified   calculation 
for di-jet hadroproduction near the back-to-back 
region.\footnote{Note 
 that interestingly in~\cite{manch}, which has 
a different point of view than  TMD, 
 Coulomb/radiative mixing terms are 
found to be 
 responsible for the 
breaking of  angular ordering in the initial-state cascade and 
 the appearance of superleading 
logarithms in di-jet cross sections with a gap in rapidity.}

A further question concerns lightcone divergences 
and the $x \to 1$ endpoint behavior. 
The singularity structure at $x \to 1$  is different  
 in the TMD case than for 
ordinary (integrated) distributions, giving divergences 
 even in dimensional regularization with 
an infrared cut-off~\cite{fhfeb07}. 
 The singularities can be understood 
in terms of gauge-invariant eikonal-line matrix elements~\cite{fhfeb07},   
and the TMD  behavior can be related 
to cusp anomalous dimensions~\cite{korchangle,chered} 
and lack of complete KLN cancellations~\cite{jcc03,korchangle,collsud}. 
 In general this  affects the precise form  of factorization and 
relation with collinear  distributions. 
 
Relevant applications are both at low energies  and at high energies.  
An important example for    current experimental programs   is 
semi-inclusive leptoproduction (\cite{anselm08,ceccopieri,muldetal}, and 
references therein), where 
infrared subtractive techniques of the type~\cite{jccfh00,jccfh01} serve for 
   TMD-factorization calculations~\cite{jiyuan}, and in particular for 
the proper 
treatment of overlapping momentum regions.\footnote{Subtraction 
techniques related to those  of~\cite{jccfh00,jccfh01}  are 
 developed in~\cite{manohstew} for soft-collinear effective theory, 
and studied in~\cite{leesterm} and~\cite{idimeh} 
in relation with standard perturbative 
methods.  See also SCET applications to  shower algorithms~\cite{bauergen},  
 TMD pdfs~\cite{chay} and jet event shapes~\cite{trottetal} for use of these 
techniques.} 
Implications for     spin asymmetries~\cite{koike} and possibly 
 exclusive reactions~\cite{bochumgpd} are being studied. 
General characterizations of TMD distributions 
will be  relevant   at  colliders for turning  
present k$_\perp$-showering generators into general-purpose 
tools to describe hadronic final states over the whole 
phase space~\cite{hj_rec,fhdistalk}.

In the next section we consider 
applications of k$_\perp$-shower generators to multi-jet final 
states~\cite{hj_ang}.    
 The main focus is  on regions where jets  are far from  back-to-back, 
 and the total  
energy is much larger than the 
transferred momenta  so that the values of $x$ are small. In this regime 
 the ambiguities   related to 
 soft   Coulomb exchange and to  lightcone divergences  are not 
expected to be crucial.   
We find that the TMD distributions,  as well  as the 
transverse-momentum dependence of short-distance matrix elements, play 
a very essential role to describe correlations in angle and momentum of 
the  jets.

\section{Multi-jet correlations }

This section presents multi-jet   results~\cite{hj_ang} of  Monte 
Carlo implementing TMD distributions according to high-energy factorization,   
and compares them with collinear-based Monte Carlo results. 

\vspace*{6 cm} 
\begin{figure}[htb]
\includegraphics{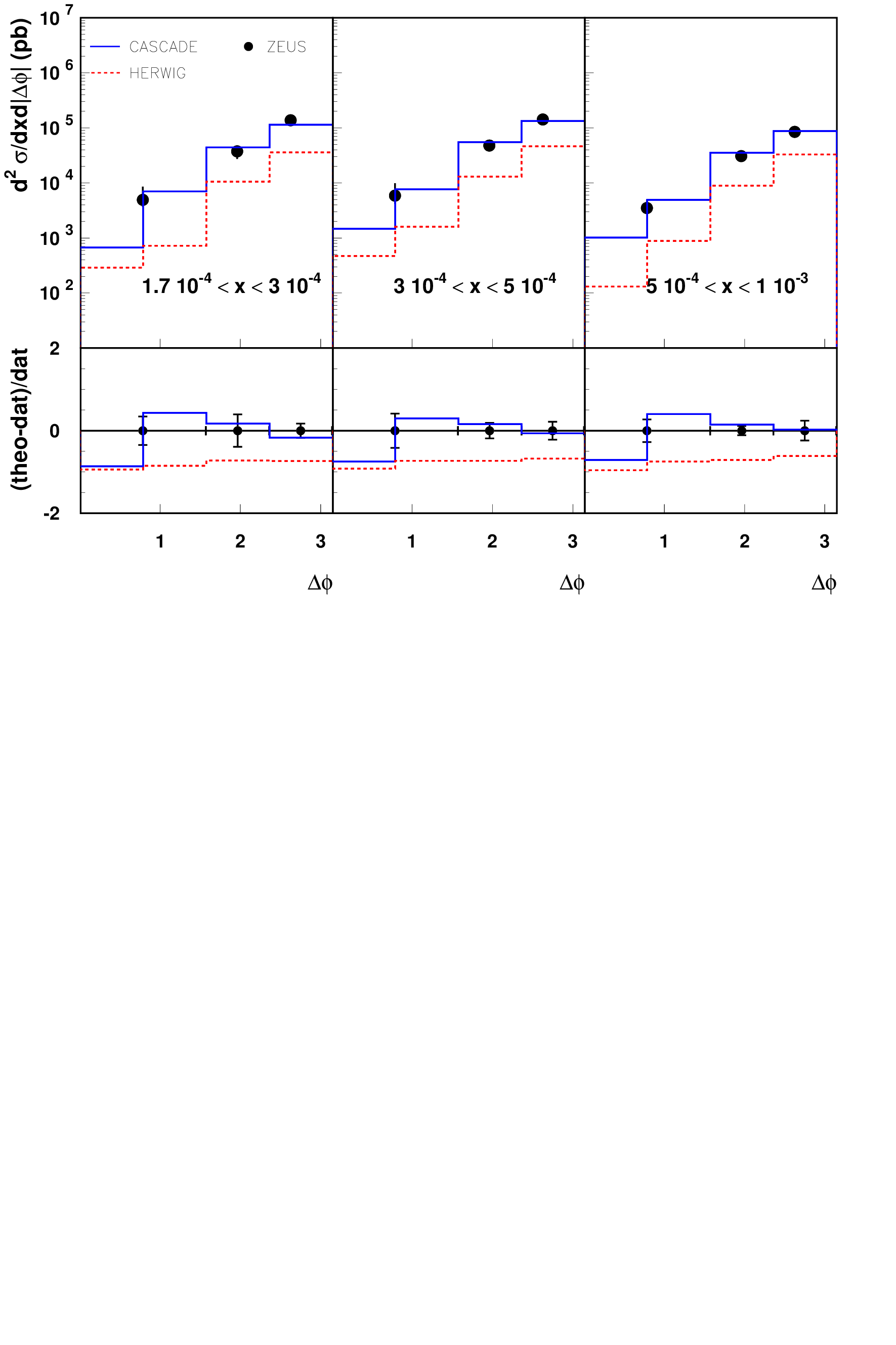}
\includegraphics{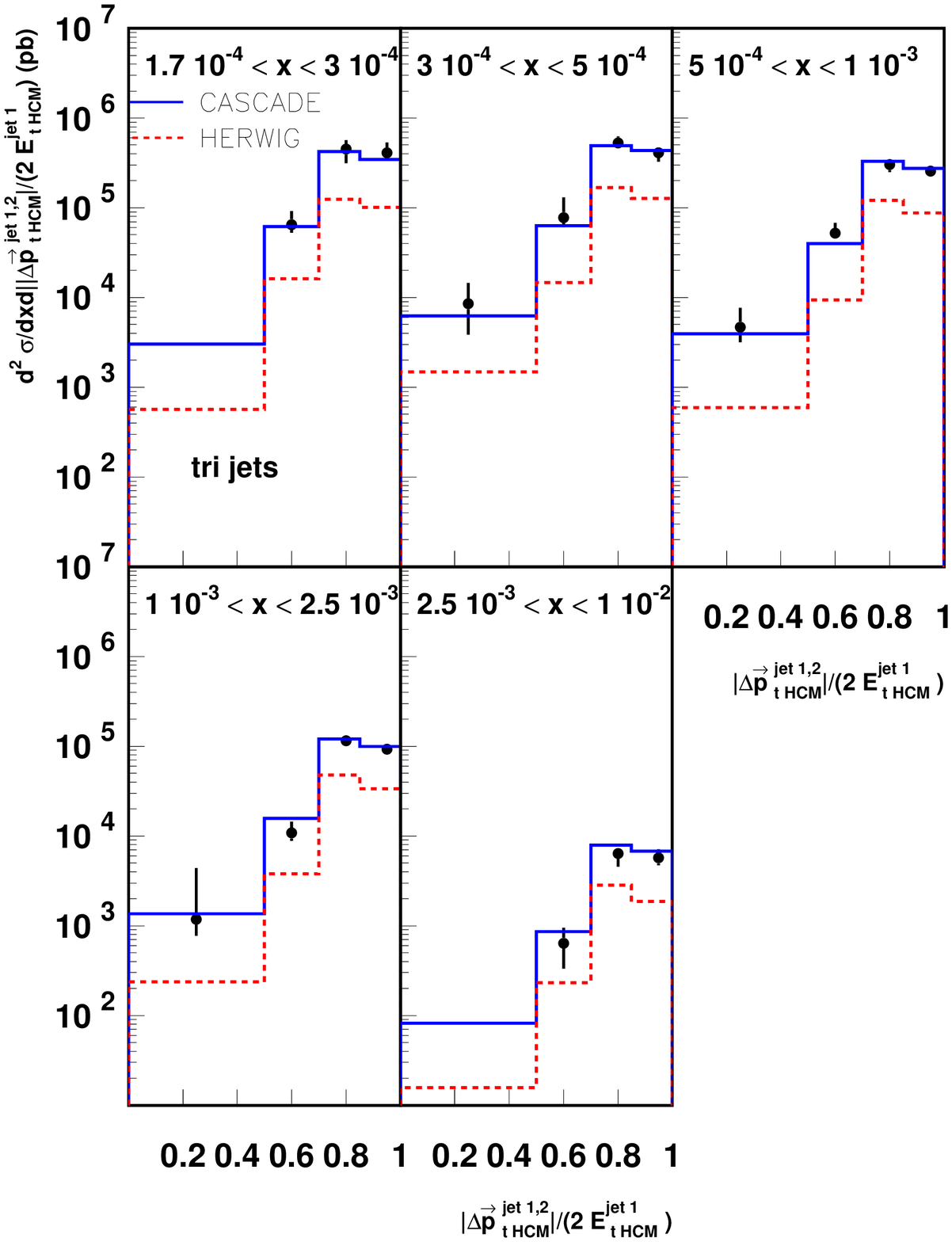}
\caption{(left) Angular  correlations and (right) 
momentum correlations~\protect\cite{hj_ang} 
in three-jet final states measured by~\protect\cite{zeus1931}, compared 
with the    \cascade\ and \herwig\ Monte Carlo results.} 
\label{fig:phipage}
\end{figure} 

In Fig.~\ref{fig:phipage} we consider three-jet production in $e p$, for which 
precise and interesting measurements have recently appeared~\cite{zeus1931}, and 
we show results for the cross section in the azimuthal separation 
$\Delta \phi$ between the two leading jets and in the transverse-momentum 
imbalance $\Delta p_t$. 
The shape of the distributions is different for \herwig\ and for the 
k$_\perp$-shower Monte Carlo  \cascade~\cite{jung02}, with the largest differences 
occurring at small $\Delta \phi$ and small $\Delta p_t$, where the 
jets are not close to  back-to-back configurations~\cite{delenda} 
and one has three hard,  well-separated jets.\footnote{Near  
$\Delta \phi \sim \pi$, on the other hand, 
  soft-gluon exchange effects such as in~\cite{vogel0708} may well 
affect the predictions.} 
By analyzing  the angular distribution of the third jet, 
Refs.~\cite{hj_rec,hj_ang}  find significant 
contributions from regions where the transverse momenta in the initial state 
shower are not ordered.  The description of the  
measurement by the k$_\perp$-shower is 
 good, whereas   the collinear-based shower  is not
sufficient to describe it.    

Note that the interpretation~\cite{hj_rec,hj_ang} of 
the angular correlation data 
in terms of corrections to collinear ordering  is 
 consistent  with the finding in~\cite{zeus1931} that 
while inclusive jet rates are reliably predicted by NLO fixed-order 
results, NLO predictions 
are affected by  large corrections to di-jet azimuthal distributions 
(going from ${\cal O} (\alpha_s^2)$ to ${\cal O} (\alpha_s^3)$) in the  
small-$\Delta \phi$ and small-$x$ region, and begin to fall below the 
data for three-jet distributions  in the smallest $\Delta \phi$ bins.

\begin{figure}[htb]
\includegraphics{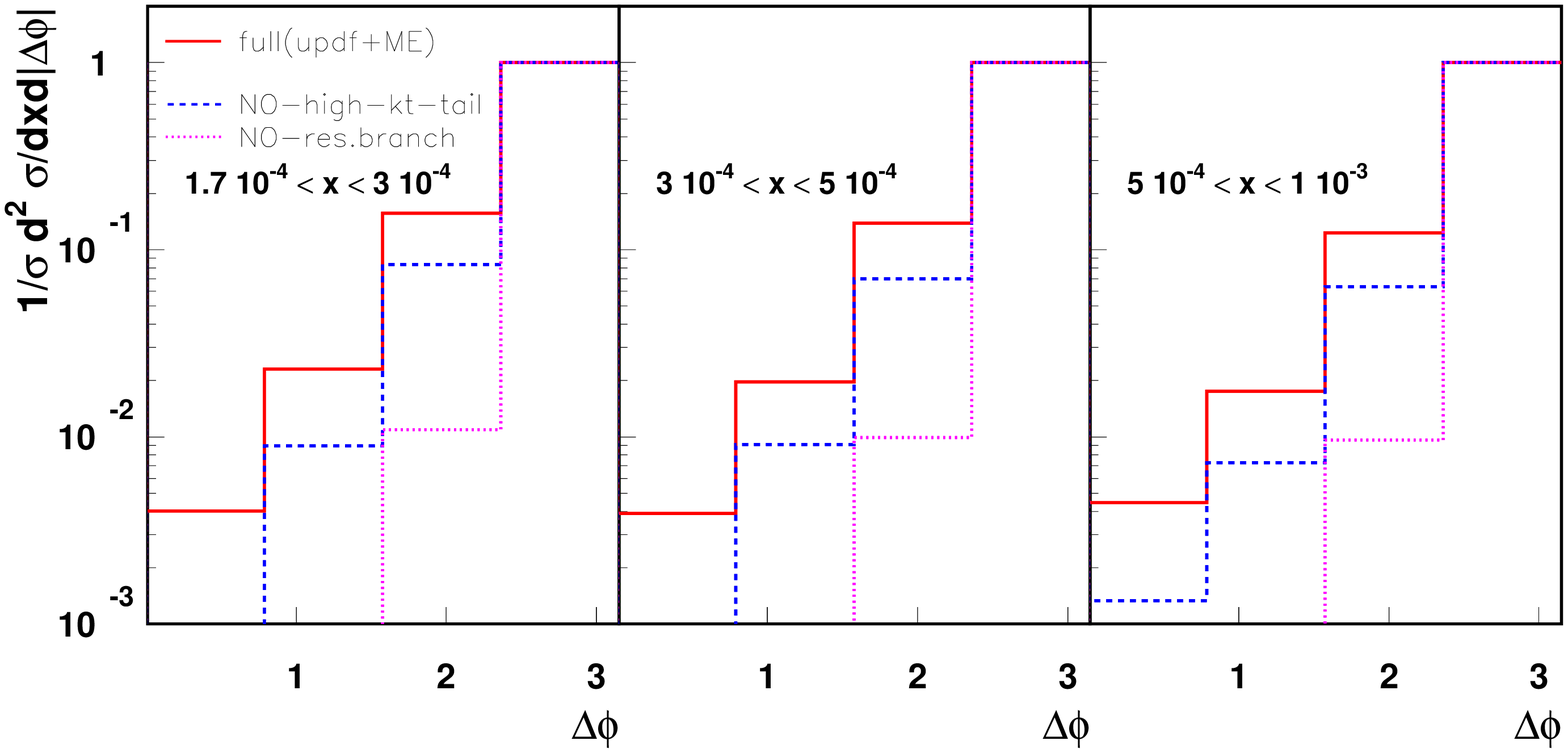}
\caption{Azimuthal distribution normalized to the 
back-to-back cross section~\protect\cite{hj_ang}: (solid red)  full result; 
(dashed blue) same TMD pdfs but no finite-k$_\perp$ correction in ME; 
(dotted violet)  TMD pdfs with no resolved branching.
}
\label{fig:ktord}
\end{figure}

It is important to realize that 
the result in Fig.~\ref{fig:phipage} receives 
contribution from the transverse-momentum 
dependence of both TMD pdfs and 
hard matrix elements. 
 Fig.~\ref{fig:ktord} 
shows different approximations to the  
azimuthal dijet distribution normalized to the 
back-to-back cross section. The solid 
red curve is   the full  result.    The 
 dashed blue curve is  obtained 
from the same TMD pdfs but  
not including the  transverse-momentum dependence of 
the hard ME.  
We see that  the 
high-k$_\perp$ component  in the hard ME~\cite{hef}  
is essential to describe 
jet correlations 
for small $\Delta \phi$.  For reference we also plot 
with the dotted (violet) curve the result     
  obtained from the  TMD pdf 
without any resolved branching, corresponding  
 to nonperturbative, predominantly 
 low-k$_\perp$ modes.

\vspace*{4 cm} 

\begin{figure}[htb]
\includegraphics{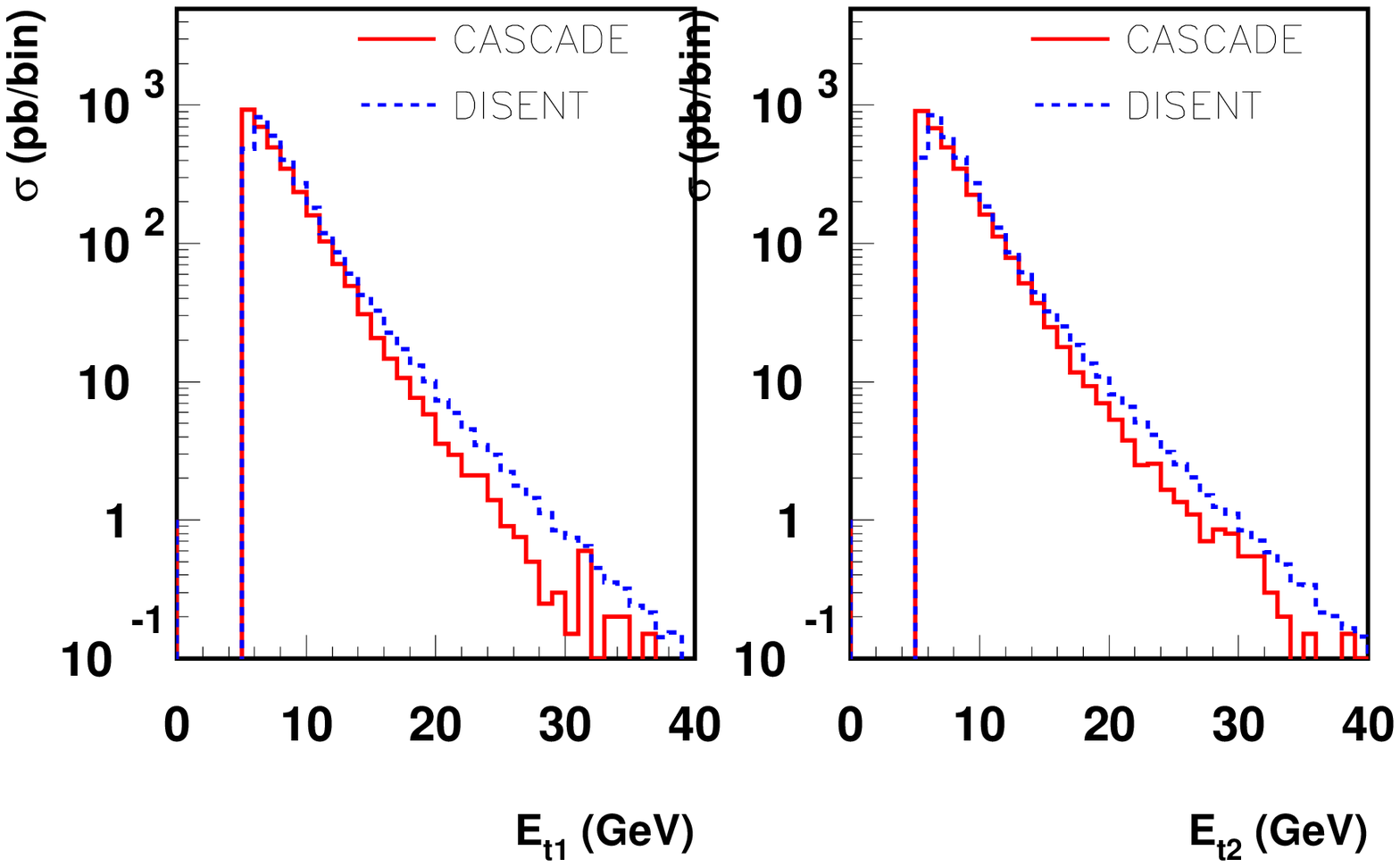}
\includegraphics{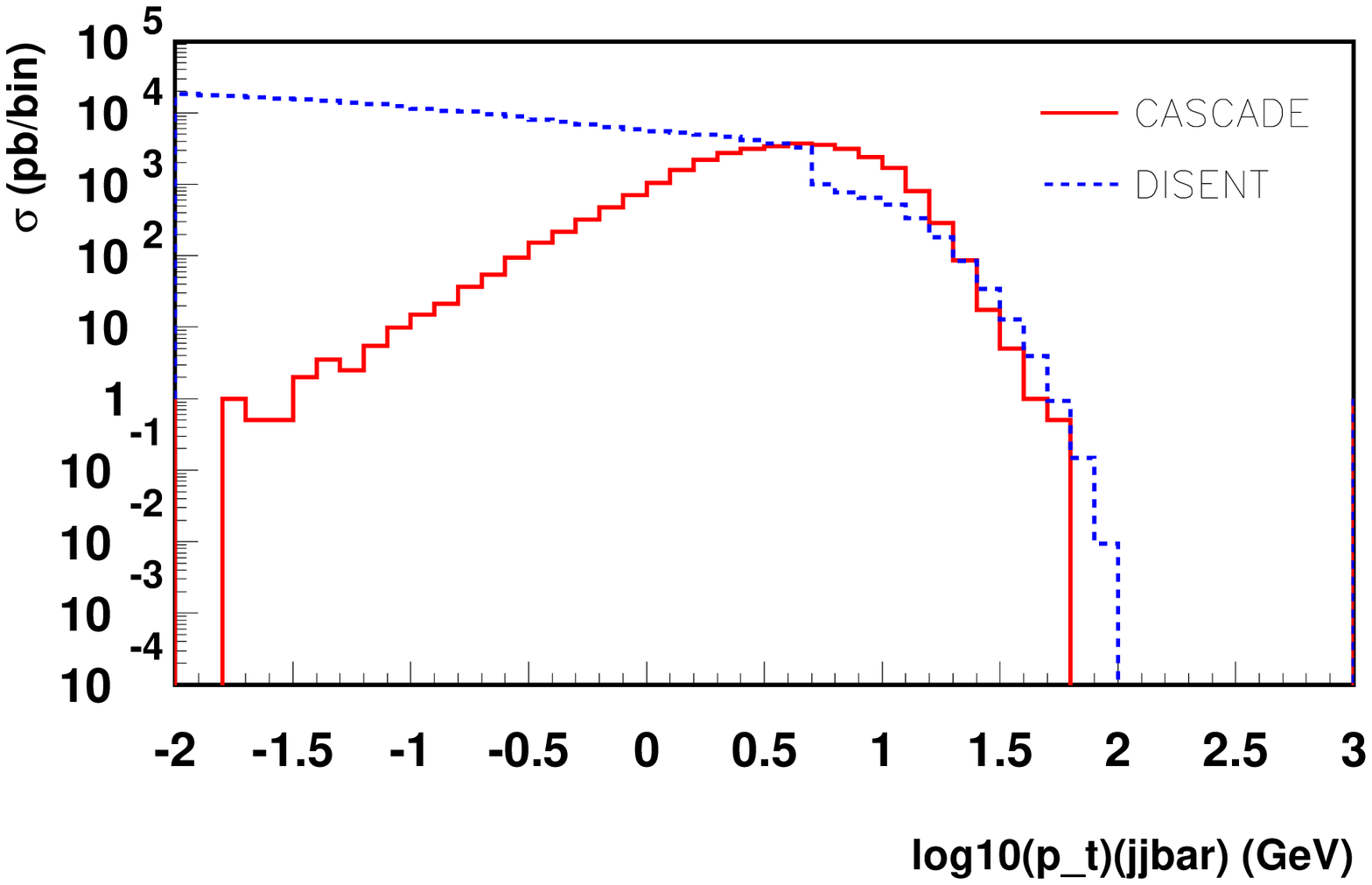}
\caption{Comparison of the k$_\perp$-shower  \cascade\ 
with the  NLO di-jet calculation \disent: (left)~distribution 
in single-jet transverse energy; (right)  distribution in the 
di-jet transverse energy.} 
\label{fig:jj}
\end{figure}

To examine more closely  the distribution in k$_\perp$ that results 
from highly off-shell  
subprocesses, in  Fig.~\ref{fig:jj} we study the jet cross section  
in transverse energy and compare the  k$_\perp$-shower with the NLO 
 result. It is noteworthy that the  large-$p_t$ part of the di-jet 
spectrum is very close for the two calculations. At low $p_t$ one 
sees the Sudakov form-factor effect in the shower result.  
Differences in the single-jet spectra are also of interest and  currently under 
study. This may be of use to relate~\cite{dasgqt}   
{\small DIS} event shapes measuring 
the transverse momentum  in the current region 
 to hadro-production $p_T$ spectra. 

\vspace*{4.6 cm} 

Let us note that 
besides  jet final states  the off-shell effects mentioned above 
are expected also for heavy mass production.  
For instance, they may affect the  
phenomenological small-$x$ broadening of W and Z $p_T$ distributions 
 emphasized by Olness~\cite{olness} (see~\cite{cpyuan1}), and their use  
as luminosity monitor~\cite{mandy}.   
Multi-scale  effects may  arise~\cite{deak} in 
the associated production of W and bottom quark pairs~\cite{mlm93}   
and in final states with Higgs~\cite{higgs02,jung-mpla}\footnote{Similar effects 
were noted~\cite{vogelhiggs} in the predictions for the 
 Higgs transverse-momentum spectrum at the LHC.} especially for 
 measurements of non-inclusive observables and correlations.

\section{Summary and prospects for  LHC final states}

We have discussed the method of  k$_\perp$-dependent Monte Carlo shower, based 
on transverse-momentum dependent (TMD), or unintegrated, parton 
distributions and matrix elements defined by high-energy factorization. 
 The theoretical basis of the method allows one to 
go to arbitrarily high transferred-momentum scales, 
thus making it suitable for the simulation of hard processes at the LHC. 

We have pointed to  developments of the approach 
toward general-purpose event generators, and  illustrated applications 
to  experimental    $e p$ data for final states with multiple hadronic jets.  
 Despite the  lower 
$ep$ energy,   the  multi-jet  kinematic region considered   is 
 characterized by  large phase space available for jet production   
and is     relevant  
 for  extrapolating to the LHC  initial-state showering effects. 

\begin{figure}[htb]
\includegraphics{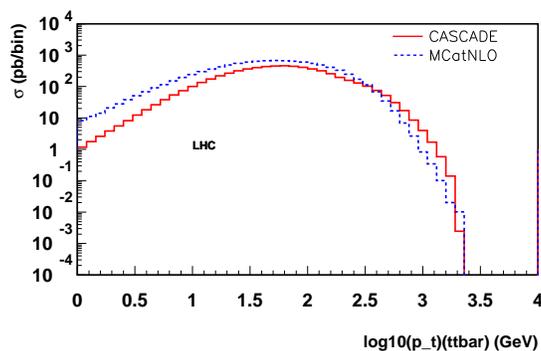}
\caption{Comparison of transverse momentum distribution of $t\bar{t}$ pairs calculated from
the k$_\perp$-shower  \cascade\ 
with the  NLO  calculation \mcatnlo\  at LHC energies.} 
\label{fig:ttbar}
\end{figure}

The multi-scale QCD effects that we are treating 
 also affect  heavy mass production at the LHC, 
including  final states with heavy bosons and heavy flavor. 
It is interesting to note that even at LHC energies the transverse momentum distribution 
of top quark pairs calculated from the k$_\perp$-shower is similar to what is obtained 
from a full 
NLO calculation (including parton showers. MC@NLO~\cite{mcatnlo}),  
where the  k$_\perp$-shower predicts an even harder spectrum,  
 Fig.~\ref{fig:ttbar}.

We conclude by observing that,  using  unintegrated parton distributions together with the 
off-shell matrix elements,  many of the sub-leading effects are properly simulated, both in $ep$ 
collisions at HERA as well as at the LHC. For $ep$ we could show that the predictions coming 
from the k$_\perp$-shower simulation \cascade\ are in good agreement with the measurements. For 
the LHC even at the large scale of $t\bar{t}$ production we observe reasonable comparison with  
results containing  full NLO effects.
\vspace*{4.6 cm} 


\begin{theacknowledgments}
  It is a pleasure to thank the organizers of the conference 
  for the  invitation and   the opportunity to 
  participate in a very interesting and enjoyable meeting. 
We wish to thank J.~Collins, M.~Mangano and J.W.~Qiu for useful discussions. 
\end{theacknowledgments}



\bibliographystyle{aipproc}   

\bibliography{sample}




\end{document}